%% file: main.tex
\documentclass{article}
\usepackage[utf8]{inputenc}
\usepackage{graphicx}
\usepackage{url}
\usepackage{braket}
\usepackage{amsmath}
\usepackage{listings}
\usepackage{bm}

\usepackage[left=2cm,right=2cm,bottom=2cm]{geometry}
\usepackage[caption = false]{subfig}
\title{Introductory Tutorial for SPSA and the Quantum Approximation Optimization Algorithm}
\author{Salonik Resch, University of Minnesota - Twin Cities\\ resc0059@umn.edu}
\date{}

\usepackage{color}

\definecolor{dkgreen}{rgb}{0,0.6,0}
\definecolor{gray}{rgb}{0.5,0.5,0.5}
\definecolor{mauve}{rgb}{0.58,0,0.82}

\begin{document}
\vspace{-10cm}
\maketitle

\begin{abstract}
    This short tutorial provides an introduction to the Quantum Approximation Optimization Algorithm (QAOA). Specifically, how to use QAOA with the Simultaneous Perturbation Stochastic Approximation (SPSA) algorithm to solve the Max-Cut problem. All steps of the algorithm are explicitly shown and no theory or complex mathematics are used. The focus is entirely on setting up a practical implementation. Fully functional examples in both R and python (using Qiskit) are provided, both using roughly 100 lines of code.
\end{abstract}
\section{Introduction}
The purpose of this tutorial is to provide an explicit description how to use the Simultaneous Perturbation Stochastic Approximation (SPSA) \cite{spall1992multivariate} in tandem Quantum Approximation Optimization Algorithm (QAOA) \cite{farhi2014quantum}. A brief overview of both is provided, before going over a concrete example. The most notable feature of this tutorial is that the optimization does not occur ``off-screen'', every step of the algorithm is visible in the examples and provided code.\footnote{With the major exception of the backend quantum mechanics simulations.} The end result is a simple implementation, but it should be helpful for getting started.

Mathematical background and theory is skipped for clarity, it is not necessary to understand in order to get started. A familiarity with quantum computation (qubits and quantum gates) is assumed. For a background on the basics of quantum computing, I highly recommend the text by Michael Loceff \cite{loceff2015course}. 

The input problem, Max-Cut, is described in Section \ref{sec:maxcut}. The QAOA circuit which can solve Max-Cut is presented in Section \ref{sec:qaoa}. The basics of SPSA are then covered in Section \ref{sec:spsa}. After a summary in Section \ref{sec:summary}, Section \ref{sec:code} goes over code which implements all components. Appendices \ref{sec:qiskit} and \ref{sec:quantumops} provide the entire code, which is also available on github \cite{code}.

\section{Max-Cut}
\label{sec:maxcut}
Before discussing the solution, let us first describe the problem. Max-Cut is a binary combinatorial optimization problem. This means we are searching for a set of binary values (which can be represented by a bitstring) which produces an optimal solution to a given cost function. If $\textbf{z}$ is a bitstring with $n$ bits
$$ \textbf{z} = z_0z_1z_2...z_{n-1}$$
we are looking for a value of $\textbf{z}$ such that a cost function $C$ is maximized
$$ argmax_\textbf{z} \;C(\textbf{z}) $$
The cost function consists of a set of \emph{clauses}, which represent constraints on the bitstring that we want to be satisfied. The input to the cost function is a bitstring (a suggested solution) and the output is an integer specifying how many of the clauses were satisfied by the bitstring. 
$$ C(\textbf{z}) = \#\;of\;clauses\;satisfied\;by\;\textbf{z}$$
For max-cut, the clauses are very simple - each clause specifies two bits in the bitstring that should be opposite values. For example, one clause might be bit 2 and bit 3 should be opposite, $z_2 != z_3$. A wide range of optimization problems can be mapped into such a format, such as 3SAT \cite{3sat}, hence using this restricted set of clauses does not significantly limit the problems we can solve. Given a bitstring, it is easy to compute how many of the clauses were satisfied. The hard part is finding the bitstring that satisfies the most clauses. As the problem size grows (the length of the bitstring), the number of possible solution bitstrings grows exponentially. Hence, a brute force search will be intractable. 
\subsection{Max-Cut Example}
\label{sec:maxcutgraph}
The input to max-cut can be described as a graph, where vertices represent the binary variables ($z_0z_1z_2...z_{n-1}$) and edges between vertices represent the constraints. If an edge connects vertices $i$ and $j$, this means that $z_i != z_j$ is one of the constraints. The problem of finding an optimal bitstring $\textbf{z} = z_0z_1z_2z_3$ is equivalent to finding an optimal 2-coloring\footnote{This is also called a graph \emph{cut}, as the coloring is equivalent to a partition and edges are ``cut'' by set boundaries when crossing from one set (color) to the other.} of the graph. Optimal means that as many edges as possible connect vertices of different colors.  

\begin{figure}[]
\centering
\subfloat[A graph with 4 vertices (variables) and 4 edges (constraints)]{\includegraphics[width=.2\textwidth]{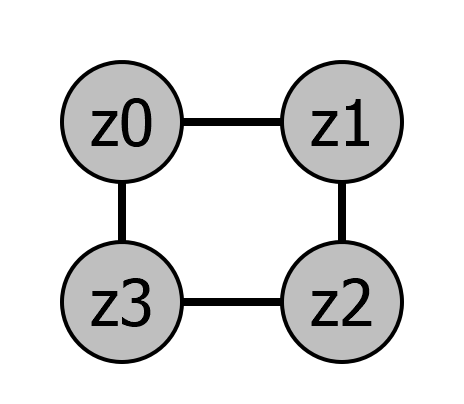}\label{fig:grapha}}
\subfloat[0000 has score 0]{\includegraphics[width=.2\textwidth]{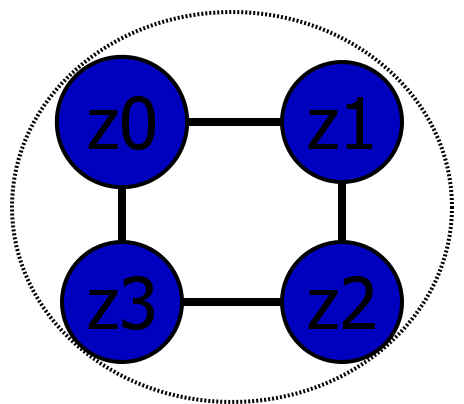}\label{fig:graphb}}
\subfloat[0011 has score 2]{\includegraphics[width=.2\textwidth]{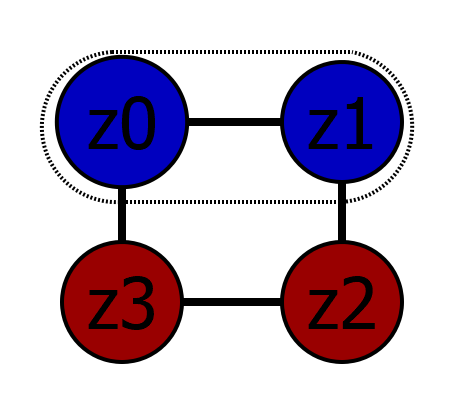}\label{fig:graphc}}
\subfloat[0101 has score 4]{\includegraphics[width=.2\textwidth]{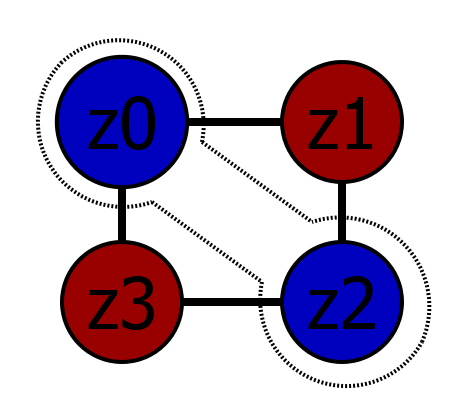}\label{fig:graphd}}
\caption{Input to max-cut can be represented by a graph where vertices are variables and edges are constraints. The problem can then be considered a graph coloring, where there are two colors (eg Blue=0, Red=1) and edges should connect vertices of different colors. The number of edges that cross color boundaries (``cut'' the set boundaries) represents the score of the coloring.}
\label{fig:graph}
%\vspace{.4cm}
\end{figure}

Let's use an example problem from Pennylane \cite{xanaduqaoa}, which has a good tutorial on QAOA. A graph with four vertices (variables) and four edges (constraints) is shown in Figure \ref{fig:grapha}. The four constraints are $z_0 != z_1$, $z_1 != z_2$, $z_2 != z_3$, and  $z_3 != z_0$. We can try to color the graph in order to satisfy as many of the constraints as possible. For example, in Figure \ref{fig:graphb} we can set $\textbf{z} = 0000$ (all blue). In this case no edges cross from blue to red, meaning that no constraints are satisfied, and the score is 0. Figures \ref{fig:graphc} and \ref{fig:graphd} show better colorings, where edges do connect vertices of different colors. In this case, all 4 constraints can be satisfied (though this is not typical). The score is also referred to as the \emph{cut}, because if a line is drawn around all vertices of the same color, it passes through (or cuts) edges of the graph. Hence the name Max-Cut.

If given a bitstring (graph coloring), we can quickly check how many constraints it satisfies (the score). This involves iterating through each of the constraints (edges), which all have the form $z_i != z_j$, and checking whether bits (vertices) $i$ and $j$ are different values (colors). We get a single point for each constraint satisfied. This has low computational complexity (linear in the number of constraints). Even for very large problems a classical computer can easily do this.

\section{Quantum Approximation Optimization Algorithm}
\label{sec:qaoa}
The quantum approximation optimization algorithm (QAOA) \cite{farhi2014quantum} is a quantum algorithm which provides an approximate solution to max-cut. At small sizes, which can be run on current quantum computers, the algorithm is much slower than alternative approaches. However, at larger scales it is believed QAOA will provide an advantage \cite{guerreschi2019qaoa}. 

QAOA is a \emph{variational} quantum algorithm, where a quantum computer and a classical computer work together to get the solution. A quantum circuit\footnote{``circuit'' is the term used for ``program'' in quantum computing literature. A quantum circuit is just a list of quantum gates (operations/instructions) to perform.} runs on the quantum computer, where the output provides a guess at the solution. The classical computer then evaluates the quality of that solution, and uses an optimization algorithm to modify the quantum circuit to try to provide a better solution (hence the use of SPSA, covered in Section \ref{sec:spsa}). The classical computer is trying to ``guide'' the quantum computer towards better solutions. This process is similar to training for machine-learning, however we don't have access to precise gradients (or any information) within the quantum computer - so we can't use tricks like backpropagation or similar machine-learning approaches. We will see that the \emph{structure} of the quantum circuit comes from the input problem, and the specific operations performed in the quantum circuit will be determined (parameterized) by \emph{classical parameters}, which are set by the classical computer.

The reason we covered the graph representation of max-cut in Section \ref{sec:maxcutgraph} is because it provides a nice visual of the conversion between the max-cut problem and a quantum circuit which solves it. We ``map'' each vertex in the graph to a single qubit. Hence, the number of qubits required is equal to the number of vertices (bits) of the input problem. The 4-bit problem in Figure \ref{fig:graph} will require 4 qubits to solve. At the end of the quantum circuit, we perform a measurement on the qubits. This produces a bitstring, one bit for each qubit. These bits (bitstring) can be interpreted as graph colorings, just as in Figure \ref{fig:graph}. So once we get the bitstring, we can quickly check how good of a score it produces. For example, if we measure $\ket{0011}$, this corresponds to the coloring in Figure \ref{fig:graphc}, and we can check to see that it produces a score of 2 on our max-cut problem.  

\begin{figure}
    \centering
    \includegraphics[scale=0.3]{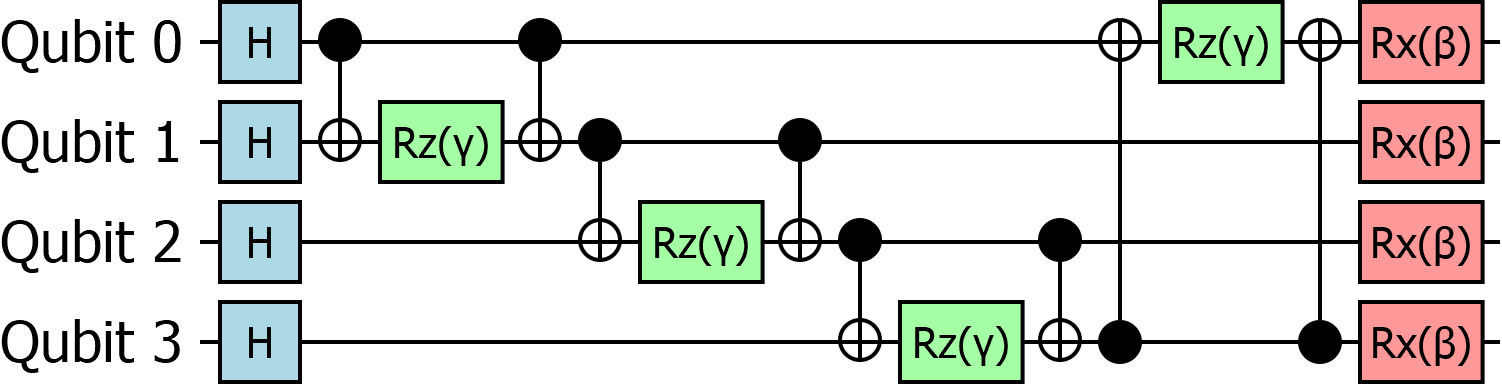}
    \caption{QAOA circuit with $p=1$. H(adamard) gates are applied to all qubits. Then a sequence of controlled z-rotations is performed. The qubits which interact via these controlled rotations is determined by the input graph, shown in Figure \ref{fig:grapha}. A controlled-z rotation is performed for each edge in the graph. Finally, x-rotations are performed on all qubits. Classical parameters $\gamma$ and $\beta$ determine the rotation angles.}
    \label{fig:ckt}
\end{figure}

Let's now describe the quantum circuit for QAOA, which is shown in Figure \ref{fig:ckt} for the input problem in Figure \ref{fig:grapha}. The qubits are initialized into the all zero state $\ket{0000}$. The first operation is a Hadamard (H) gate on all qubits, which puts them into an even superposition of all states. 

\begin{figure}
    \centering
    \includegraphics[scale=0.3]{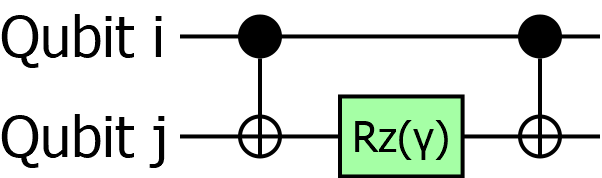}
    \caption{A controlled phase operation between qubit $i$ and qubit $j$. Two CNOT gates and a single-qubit z-rotation is used. The angle of the z-rotation is set by $\gamma$, which is a classical parameter.}
    \label{fig:controlledz}
\end{figure}

We then do a sequence of controlled-phase operations. A single controlled phase operation can be performed with three sequential operations. Say we are doing a controlled-phase operation between qubit $i$ and qubit $j$. First, a CNOT gate is performed with qubit $i$ as the control and qubit $j$ as the target. Then a single-qubit z-rotation is performed on qubit $j$. Finally, another CNOT is performed, again with qubit $i$ as the control and qubit $j$ as the target. This is shown in Figure \ref{fig:controlledz}. Note that the z-rotation is parameterized by a parameter $\gamma$, which sets the angle of rotation

The sequence of controlled-phase operations that we need to perform is determined by the input graph. For every edge in the graph, we perform a single controlled-phase rotation. If there is an edge between vertices $i$ and $j$, we perform a controlled phase rotation between qubit $i$ and qubit $j$. However, it does not matter which qubit is the control and which is the target. Additionally, the order of the controlled-rotations can permuted \cite{alam2020circuit}, any order is fine. The same $\gamma$ is used for all controlled phase rotations. $\gamma$ is set by the classical computer, which is covered in Section \ref{sec:spsa}. 

Once all the controlled phase operations have been completed, single qubit x-rotations are performed on all qubits. The x-rotations are parameterized by $\beta$ (also set by the classical computer), which is the same for all qubits. After the circuit is finished, a measurement is performed on the qubits. The measurement will destroy the quantum state and will produce a single classical bit for each qubit in the circuit. This bitstring can be interpreted as a graph coloring of the input problem.

We can increase the depth of the QAOA circuit (length of the QAOA program) in order to achieve a better answer. A hyperparameter $p$ sets the depth. If $p=1$, the QAOA circuit is exactly as shown in Figure \ref{fig:ckt}. If $p>1$, we repeat the controlled phase operations and x-rotations $p$ times (the H gates are not repeated). Each stage will use different values for the parameters $\gamma$ and $\beta$. Hence, there will be a total of $2p$ classical parameters which determine the QAOA circuit, $\pmb{\gamma} = \gamma_0\gamma_1... \gamma_{p-1}$ and $\pmb{\beta} = \beta_0\beta_1... \beta_{p-1}$. A higher $p$ will produce better results in absence of noise. However, as modern quantum computers are very noisy, a higher $p$ can actually produce worse results, as the susceptibility to noise increases with the length of the program.

It should be noted that QAOA circuit will typically not produce the same output every time, even if we perform the exact same operations with the same parameters. Each time we run the quantum circuit and perform the measurement we're likely to get different bitstrings - the circuit is effectively creating a \emph{distribution} of output bitstrings. Hence, we will need to run the circuit many times to get a good estimate on what the distribution is. We will be interested in the \emph{expectation value} of the distribution, which is simply the average ``score'' of all the bitstrings in the distribution. As we run QAOA, we will try to find values for $\bm \gamma$ and $\bm \beta$ that produce the largest expectation value - covered in Section \ref{sec:spsa}.

\section{Simultaneous Perturbation Stochastic Approximation}
\label{sec:spsa}
\textbf{Notation:} I use a short-hand representation of vector operations. I will use bold for vectors and regular print for scalers. For example, if $\pmb x = [1,2,3,4,5]$, a vector (element-wise) operation can be written as
\begin{equation}
    \pmb y = 1/\pmb x = \left [ \frac 1 1 ,\frac 1 2 ,\frac 1 3 ,\frac 1 4 ,\frac 1 5 \right ]
    \label{eq:vectors}
\end{equation}

Before discussing how we optimize the QAOA circuit to produce better results, it's useful to cover some background on the classical optimizer. The Simultaneous Perturbation Stochastic Approximation (SPSA) algorithm \cite{spall1992multivariate} is commonly used to optimize QAOA. SPSA finds widespread use in industry due to its many benefits. It's simple, has resilience to noise, and remains efficient even for very large optimization problems. It's a good choice whenever one does not have access to the inner workings of the function they are trying to optimize (we don't on quantum computers as they are obscured within the quantum mechanics). 
Say we have a function $F$ that we'd like to optimize. $F$ is parameterized by a vector of $m$ parameters, $\bm \Theta = \theta_0\theta_1...\theta_{m-1}$. 
We'd like to maximize $F(\bm \Theta)$. 

We start by randomly initializing $\bm \Theta$ by assigning a random number to each $\theta$. We then generate a random perturbation vector $\bm \Delta = \delta_0\delta_1...\delta_{m-1}$. Each $\delta$ has a magnitude of $c$, but can be positive or negative (what $c$ is will be covered in a bit). So $\bm \Delta$ is a vector of Bernoulli random variables (which can be $\pm 1$) multiplied by a constant $c$. From this perturbation vector we create two new versions of our parameter vector
\begin{equation}
    \begin{aligned}
    \bm \Theta_+ &= \bm \Theta + \bm \Delta &= [\theta_0+\delta_0, \theta_1+\delta_1,...,\theta_{m-1}+\delta_{m-1}]\\
    \bm \Theta_- &= \bm \Theta - \bm \Delta &= [\theta_0-\delta_0, \theta_1-\delta_1,...,\theta_{m-1}-\delta_{m-1}]
    \end{aligned}  
\end{equation}
We have one parameter vector with the perturbations added and one parameter vector with the perturbations subtracted. This will allow us to estimate the gradients ($\pmb g$) of the perturbation vector by performing two evaluations of our objective function
\begin{equation}
    \pmb g = \frac{ F(\bm \Theta_+) - F(\bm \Theta_-) } { 2c \bm \Delta }
\end{equation}

$ F(\bm \Theta_+)$ and $F(\bm \Theta_-)$ are the outputs of the objective function with the two parameter configurations, and they are both scalers. c is also a scaler. $\pmb \Delta$ is a vector of length $m$, so $\bm g$ is also a vector of length $m$.\footnote{Note that $\frac{ F(\bm \Theta_+) - F(\bm \Theta_-) } { 2c  }$ is a scaler. So $\pmb g$ is this scaler divided by the vector $\pmb \Delta$, just like in Equation \ref{eq:vectors}} If $F(\bm \Theta_+) > F(\bm \Theta_-)$, we'll move the parameters towards $\bm \Theta_+$. Otherwise, we'll move the parameters towards $\bm \Theta_-$. This is done with our parameter update function.
\begin{equation}
    \bm \Theta = \bm \Theta + a\times \pmb g
\end{equation}
We update the add  a fraction of the gradients to each of our parameters. The parameter $a$ (a scaler) determines how large this update is, and it is analogous to the learning rate in machine learning. The update of the parameter vector $\bm \Theta$ concludes one optimization iteration. Many iterations will be required to find the optimal parameters. A new perturbation vector $\bm \Delta$ will be generated for each iteration.

In the equations above, $a$ and $c$ are scalers which determine how much we perturb our parameters ($c$) and how much we update the parameters based on the gradients ($a$). We'll need to reduce $a$ and $c$ over time (over the optimization iterations) to allow the problem to converge to a final result. For optimization iteration $i$, we'll typically use something like
\begin{equation}
    \begin{aligned}
    a &= a_{start} / (i+1)^{decay} \\
    c &= c_{start} / (i+1)^{decay} \\
    \end{aligned}
\end{equation}
where $a_{start}$ and $c_{start}$ are the initial values, $i$ is the index of the optimization iteration, and $decay$ is the rate at which the parameters decay.\footnote{More complex variations of these sequences exist \cite{spsasite}, such as having different decay rates for each sequence. However, I will be sticking with this relatively simple variant which works sufficiently well.} Each of the parameters (their initial values and the decay rates) can be finely tuned, and the optimal values for each will depend on the specific optimization problem. If you want to be smart, one can use the gradient found on the the first optimization iteration to set $a$ or $c$ appropriately \cite{kandala2017hardware}. However, in the examples provided later, I simply tried multiple values until I found a few that worked well. 

We can see where the name of SPSA comes from, we are doing a \emph{simultaneous perturbation} of all our parameters. This is in contrast to gradient descent, where we would perturb each parameter individually. Doing it simultaneously means we only have to perform two evaluations of our objective function for each optimization iteration, regardless of the number of parameters.

\subsection{SPSA applied to QAOA}
\begin{figure}
    \centering
    \includegraphics[scale=0.45]{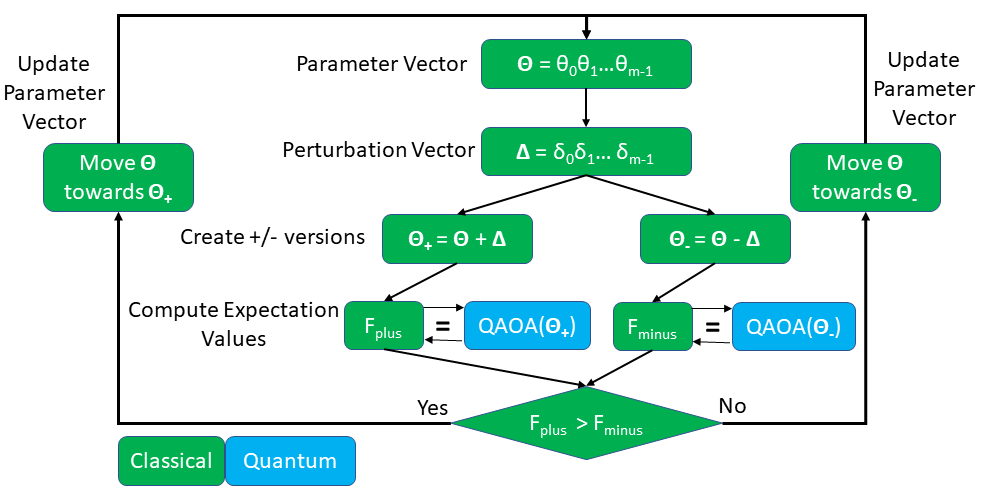}
    \caption{The process of using SPSA to train QAOA. Green blocks are performed on the classical computer, Blue blocks are performed on the quantum computer}
    \label{fig:flow}
\end{figure}

When SPSA is used with QAOA, the function $F$ produces the expectation value coming from the QAOA circuit and $\bm \Theta$ is the $\pmb \gamma$ and $\pmb \beta$ parameters that set the rotation angles of the quantum gates. SPSA is running on the classical computer, choosing values for $\pmb \gamma$ and $\pmb \beta$. It then sends the parameters to the quantum computer, which will run the QAOA circuit and generate the expectation value. Based on the resulting expectation values, SPSA will update $\pmb \gamma$ and $\pmb \beta$. Hence, SPSA will be evaluating the performance of two slightly different QAOA circuits (which have slightly varying $\gamma$ and $\beta$) and it will move the parameters towards which circuit performs better. This process is shown pictorially in Figure \ref{fig:flow}.

\section{High Level (Re)View}
\label{sec:summary}
Before getting into the specifics of the implementation, it is useful to look at the situation from a high level to see how all the working parts interact. 

The \textbf{input graph} represents the optimization problem we'd like to solve. The binary variables are represented by vertices and the constraints are represented by edges. We want to find graph colorings (bitstrings) which satisfy the problem optimally. We want vertices connected by edges to be different colors.

A \textbf{QAOA circuit} can be constructed deterministically based on the input graph. It specifies the quantum operations we need to perform. H gates are performed at the beginning. For each stage, controlled phase rotations are performed for each edge and single qubit x rotations are performed on all qubits. The rotation angles are determined by the parameters $\gamma$ and $\beta$, which are unknown at the start. The number of stages is determined by the hyperparamter $p$, which can be set arbitrarily. 

\textbf{SPSA} is going to search for optimal values of $\pmb \gamma$ and $\pmb \beta$ which will enable QAOA to solve the problem. It (typically) starts by setting setting $\pmb \gamma$ and $\pmb \beta$ randomly. Then it it will perturb them by some amount, creating two different versions. SPSA will call the QAOA circuit with both sets of parameters and compare the relative quality of both. It will then update the original parameters in the direction of which set performed better.

\section{Specific Implementations}
\label{sec:code}
In this section I will provide a walk-through of code which implements SPSA and QAOA to solve a given max-cut problem, a python (Qiskit) version in Section \ref{sec:explainqiskit} and an R (QuantumOps) version in Sections \ref{sec:explainquantumops}. The code is available for download from github \cite{code}.

\subsection{Pre-Requisites}
The R implementation requires the ``QuantumOps'' package \cite{quantumops} which is available on the CRAN repository. QuantumOps is just a library that will supply the matrices which represent the quantum operations that we need to perform. The python version requires IBM's Qiskit \cite{cross2018ibm}.

\subsection{Inputs}
The problem requires the following inputs 
\begin{itemize}
    \item \textbf{n}: The number vertices of input graph. Also is the length of the graph coloring bitstrings and the number of qubits required
    \item \textbf{edges}: The edges between the vertices representing the constraints. In R, this is a two-column matrix where source vertices are in column 1 and destination vertices are in column 2. In python this is a list of lists
    \item \textbf{p}: The depth of the QAOA circuit we want to use
    \item \textbf{nIterations}: The number of optimization iterations to run SPSA for. Each iteration consists of two calls to QAOA
    \item \textbf{nSamples}: The number of samples (measurements/bitstrings) to collect from the QAOA circuit for each optimization iteration. This corresponds to the number of times the QAOA circuit needs to be run. 
    \item \textbf{a\_start}: The starting value for $a$
    \item \textbf{c\_start}: The starting value for $c$
    \item \textbf{decay}: The decay factor for $a$ and $c$.
\end{itemize}

\subsection{python implementation (Qiskit)}
\label{sec:explainqiskit}
\input{qiskit}

\subsection{R implementation (QuantumOps)}
\label{sec:explainquantumops}
\input{quantumops}

\appendix
\input{appendixqiskit}
\input{appendixquantumops}

\bibliographystyle{acm}
\bibliography{ref}
\end{document}

%% file: qiskit.tex
\subsubsection{Computing Expectation Value}
We need to be able to compute the score of a bitstring (graph coloring) - how well it solves the optimization problem. As described in Section \ref{sec:maxcutgraph}, this involves checking each edge to see if it connects vertices (bits) of different colors (values). When simulating a quantum circuit, Qiskit returns a python dictionary where the keys are the measurement values and the values are the number of times that measurement was observed. Given this dictionary, \textbf{counts}, and the \textbf{edges} of the input graph we can find the expectation value of the distribution of outputs. We iterate over all the observed bitstrings, then check the bits (vertices) on both ends of all the edges. If the vertices (bits) are different colors (values), we increment the score. The average score of all the bitstrings is the expectation value.
\begin{lstlisting}
#Compute all scores for a set of edges
def computeExpectationValue(counts,edges):
    totalScore = 0
    totalSamples = 0
    #For each bitstring measured (keys in counts dictionary)
    for bitstring in counts.keys():
        score = 0  #Score for this bitstring
        #For each edge
        for j in range(len(edges)):
            #If vertices (bits) on both ends of edge are different
            if( bitstring[edges[j][0]] != bitstring[edges[j][1]] ):
                score += 1  #Increment score
        totalScore += score * counts[bitstring]  #Multiply score times the # of times it was observed
        totalSamples += counts[bitstring]        #Keep track of the number of measurements (samples)
    return(totalScore/totalSamples)
\end{lstlisting}

\subsubsection{QAOA Circuit}
We need to create the QAOA circuit that runs on the quantum computer. The circuit depends on both on the input graph and the classical parameters. The following function creates a quantum circuit as described in Section \ref{sec:qaoa}. It starts with Hadamards on all qubits. Then for $p$ iterations, we apply controlled z-rotations for each edge of the input graph (\textbf{graph}) and then do x-rotations on all qubits. \textbf{betas} and \textbf{gammas} are the $\pmb \beta$ and $\pmb \gamma$ parameter vectors that will be supplied by the classical computer.
\begin{lstlisting}
#Create quantum circuit for QAOA from edges and parmeters
def QAOA(nQubits,edges,p,betas,gammas):  
    #Define quantum and classical registers
    qr = QuantumRegister(nQubits)
    cr = ClassicalRegister(nQubits)
    circuit = QuantumCircuit(qr,cr)
    
    #Initial Hadamards
    for q in range(nQubits):
        circuit.h(q)        
    #For the number of specified iterations
    for P in range(p):
        #Controlled phase rotations
        #For each edge
        for j in range(len(edges)):
            #First CNOT from source qubit to destination qubit
            circuit.cx(edges[j][0],edges[j][1])           
            #Rz on destination qubit
            circuit.rz(phi=gammas[P],qubit=edges[j][1])           
            #Second CNOT from source qubit to destination qubit
            circuit.cx(edges[j][0],edges[j][1])
        
        #X rotations    
        for q in range(nQubits):
            circuit.rx(theta=2*betas[P],qubit=q)   
    circuit.measure(qr,cr)
    return circuit
\end{lstlisting}

\subsubsection{Auxiliary Functions}
Some additional functions keep the code clear and clean. The following function takes an input quantum circuit, calls the simulator, and returns the counts dictionary (measurement results). \textbf{shots} is the number of samples to take (repetitions and measurements of the quantum circuit).
\begin{lstlisting}
#Run the circuit and return counts
def runCKT(circuit,shots=10000):
    simulator = Aer.get_backend('qasm_simulator')
    counts = execute(circuit,backend=simulator,shots=shots).result().get_counts(circuit)  
    return counts
\end{lstlisting}
The following function takes an input circuit, gets the counts by calling \emph{runCKT} and then finds the expectation value by calling \emph{computeExpectationValue}.
\begin{lstlisting}
#Run a circuit and get expectation value
def ExpectationValue(circuit,edges,nSamples=10000):
    #Run circuit and collect counts
    counts = runCKT(circuit=circuit,shots=nSamples)
    #Get the score of the counts
    score = computeExpectationValue(counts,edges)
    return(score)
\end{lstlisting}
This function conveniently produces a bernoulli random variable for the perturbation vectors
\begin{lstlisting}
#Return +1 or -1 with equal probability
def m1p1():
    return random.randrange(2)*2 - 1
\end{lstlisting}

\subsubsection{SPSA}
We now have all the parts to run SPSA to optimize QAOA. The following funtions searches for optimal values of $\pmb \gamma$ and $\pmb \beta$. It starts by initializing the $a$ and $c$ sequences we require, which decay over time. It then initializes $\pmb \gamma$ and $\pmb \beta$ randomly, but near zero. It then iterates through the optimization process many times. Each iteration does the following
\begin{enumerate}
    \item Set perturbation vectors \textbf{Delta\_gammas} and \textbf{Delta\_betas}, whose values are all $\pm c$
    \item Creates two versions of the QAOA circuit, one with the perturbations added and one with them subtracted 
    \item Finds the expectation value for each circuit
    \item Computes the gradients \textbf{g\_gammas} and \textbf{g\_betas} from the expectation values and the perturbations
    \item Updates the parameters depending on the gradients and learning rate $a$
\end{enumerate}
\begin{lstlisting}
def SPSAforQAOA(n,edges,p,nIterations,nSamples,a_start,c_start,decay):
    #Initiate
    a = []; c = [];
    for i in range(1,nIterations+1):
        a.append( a_start / (i ** decay) )
        c.append( c_start / (i ** decay) )
    for i in range(nIterations):
        if( c[i] < 0.01 ):
            c[i] = 0.01
            
    #Initiate gamma,beta
    gammas = []
    betas  = []
    for P in range(p):
        gammas.append( random.uniform(-.1,.1) )
        betas.append(  random.uniform(-.1,.1) )
    
    #Run iterations of SPSA
    for i in range(nIterations):
        #Randomly perturb gammas,betas by c[i]
        Delta_gammas = []; gammas_plus = []; gammas_minus = [];
        Delta_betas  = []; betas_plus  = []; betas_minus  = [];
        for P in range(p):
            #Generate perturbation vectors of bernoulli variables with magnitude c[i]
            Delta_gammas.append( m1p1() * c[i] )
            Delta_betas.append(  m1p1() * c[i] )
            #Create +/- versions of the parameters
            gammas_plus.append( gammas[P] + Delta_gammas[P])
            gammas_minus.append(gammas[P] - Delta_gammas[P])
            betas_plus.append( betas[P] + Delta_betas[P])
            betas_minus.append(betas[P] - Delta_betas[P])
         
        #Get the circuits for the +/- versions
        pCircuit = QAOA(nQubits=n,edges=edges,p=p,betas=betas_plus,gammas=gammas_plus)
        mCircuit = QAOA(nQubits=n,edges=edges,p=p,betas=betas_minus,gammas=gammas_minus)
        #Run the +/- versions
        Fplus =  ExpectationValue(circuit=pCircuit,edges=edges,nSamples=nSamples)  
        Fminus = ExpectationValue(circuit=mCircuit,edges=edges,nSamples=nSamples)  
        
        #Compute estimated gradients 
        g_gammas = []; g_betas = [];
        for P in range(p):
            g_gammas.append( (Fplus - Fminus) / (2*Delta_gammas[P]) )
            g_betas.append(  (Fplus - Fminus) / (2*Delta_betas[P])  )
            
        #Update the parameters
        for P in range(p):
            gammas[P] = gammas[P] + a[i]*g_gammas[P]
            betas[P]  = betas[P]  + a[i]*g_betas[P]
        
        #Report progress
        print('Iteration:',i,'Exp(+):',Fplus,'Exp(-):',Fminus)
\end{lstlisting}

The code can be run out of the box with the given example
\begin{lstlisting}
def example():
    #4 qubits
    n = 4
    #Edges of the maxcut problem
    edges = [ [0,1] , [1,2] , [2,3] , [3,0] ]
    #p=2 is sufficient for this problem
    p = 2
    #A sufficient number of optimization iterations to solve problem
    nIterations = 100
    #Typically need quite a few samples (measurements of quantum circuit) per iteration to 
    nSamples = 10000
    #Heuristically chosen a and c
    a_start = 0.25
    c_start = 0.25
    decay = 0.5
    SPSAforQAOA(n=n,edges=edges,p=p,nIterations=nIterations,nSamples=nSamples,
    a_start=a_start,c_start=c_start,decay=decay)
\end{lstlisting}

%% file: quantumops.tex
\subsubsection{Computing Score}
As mentioned, computing the score (the number of constraints satisfied by a bitstring) is a classically efficient process. We need a function to perform this action, so we can evaluate the quality of the bitstrings coming out of QAOA. In addition to the bitstring to evaluate, we need to know the edges of the graph to compute how many of the edges were cut by the bitstring (graph coloring). We can compute the score by checking each edge in the graph and seeing whether it connects vertices (bits) of opposite colors (values). If the edge does connect vertices of opposite colors, we increment the score by 1. Note that the edges are represented as a 2-column matrix, where rows are edges and the two columns are the source and destination vertices. dim(edges)[1] is the number of rows in the matrix. 

\begin{lstlisting}
#classically check the score (cut) of a bitstring
computeScore <- function(bitstring,edges){
    score <- 0
    for(e in 1:dim(edges)[1]){  #For each edge  
        #If bits on either end of the edge are different
        if(bitstring[edges[e,1]+1] != bitstring[edges[e,2]+1])  #R indices start at 1, offset index by +1
            score <- score + 1                             #Score increases by 1
    }
    score #return score
}
\end{lstlisting}

\subsubsection{Quantum Circuit}
We need to create a quantum circuit for QAOA which depends both on the input problem and the classical parameters. In QuantumOps quantum circuits are represented by matrices. We build the matrix by going through the quantum circuit and adding operations as we go. Say we are creating a matrix $m$ which representations two consecutive operations $op1$ and $op2$, also represented by matrices. We can construct $m$ with
\begin{lstlisting}
m <- op1
m <- op2 %*% m 
\end{lstlisting}
\%*\% is matrix multiplication in R. The following function creates a QAOA circuit. It uses QuantumOps functions \textit{many}, \textit{controlled}, and \textit{single}. \textit{many} creates a tensor product of many identical operations, used to create the idential H and Rx gates on all qubits. \textit{controlled} creates controlled versions of gates, a controlled-x gate is a CNOT gate. \textit{single} creates a single gate which is in a tensor product with many \textit{Idle} gates, the relevant gate is applied to a single qubit and the rest of the qubits are idle.

\begin{lstlisting}
#Create QAOA circuit (represented by a matrix) from edges and parameters
QAOA <- function(nQubits,edges,p,gammas,betas){
  #Initial Hadamard gates
  m <- many(gate=H(),n=nQubits)
  #For each stage
  for(P in 1:p){
    #Phase separation for each edge
    for(e in 1:dim(edges)[1]){
      #CNOT from source to destination
      m <- controlled(gate=X(),n=nQubits,cQubits=edges[e,1],tQubit=edges[e,2]) %*% m 
      #Rz on destination
      m <- single(gate=Rz(gammas[P]),n=nQubits,t=edges[e,2]) %*% m  
      #CNOT from source to destination
      m <- controlled(gate=X(),n=nQubits,cQubits=edges[e,1],tQubit=edges[e,2]) %*% m  
    }
    #Rx on all qubits
    m <- many(gate=Rx(2*betas[P]),n=nQubits) %*% m
  }
  m #Return the matrix
}
\end{lstlisting}

\subsubsection{Expectation Value}
We are interested in the expectation value produced by the QAOA circuit for a given set of $\gamma$ and $\beta$. In order to get an accurate estimate of the expectation value we'll need to run the circuit many times. The following function takes in the relevant parameters and reports back the expectation value. It starts by constructing the required quantum circuit, with a call to \emph{QAOA}. Then it creates the $\ket{00...0}$ quantum state with as many qubits as required with a call to \emph{intket}. The circuit is then applied to the quantum state. In experiment, the quantum state would then be destroyed by a measurement and we would have to re-perform the circuit many times to get all the measurements. With simulation we can cheat and directly look at the measurement probabilities, which is provided by calling \emph{probs} on the quantum state. We can then sample integers from that distribution, emulating many repeated measurements on the quantum state. We then need to evaluate each of the measurements (bistrings) on how well they solve the optimization problem. We return the average score, the expectation value.

\begin{lstlisting}
ExpectationValue <- function(n,edges,p,gammas,betas,nSamples){
    #Generate the QAOA quantum circuit 
    circuit <- QAOA(nQubits=n,edges=edges,p=p,gammas=gammas,betas=betas)
    #Initialze a qubit register in the |00...0> state
    qr <- intket(x=0,n=n)
    #Apply the circuit (represented by a matrix) to the quantum register (represented by a vector)
    qr <- circuit %*% qr
    #Simulate the measurements of the quantum state 
    #each measurement would correspond to individual run in experiment
    #Sample from 0 to n-1 with probabilities from the quantum state
    measurements <- sample(x=0:(2^n-1),size=nSamples,prob=probs(qr),replace=TRUE) 
    
    #Test the score of the measurements
    scoreSum <- 0     #Add up all the scores from all measurments
    for(i in 1:nSamples)  #convert integer result to binary bitstring before parsing
        scoreSum <- scoreSum + computeScore(bitstring=convert_dec2bin(measurements[i],len=n),edges=edges)
    expectationValue <- scoreSum/nSamples #Average them
    
    return(expectationValue)
}
\end{lstlisting}

\subsubsection{SPSA}
All the parts now exist in order to effectively use SPSA to optimize QAOA. The following function searches for optimal values of $\gamma$ and $\beta$. It starts by initializing the $a$ and $c$ sequences we require, which decay over time. It then initializes $\gamma$ and $\beta$ randomly. It then iterates through the optimization process many times which consists of
\begin{enumerate}
    \item Set perturbation vectors \textbf{dGammas} and \textbf{dBetas}, whose values are all $\pm c$
    \item Finds the expectation value for the two versions of QAOA, one with the perturbations added and one with them subtracted
    \item Computes the gradients \textbf{gGammas} and \textbf{gBetas} from the expectation values and the perturbations
    \item Updates the parameters depending on the gradients and learning rate $a$
\end{enumerate}
Note that R conveniently does element-wise multiplation/addition when two vectors are supplied as arguments
\begin{lstlisting}
#Run Experiment
SPSAforQAOA <- function(n,edges,p,nIterations,nSamples,a_start,c_start,decay=2){
    #Make learning rate and perturbation sequences
    a <- a_start / (1:nIterations)^decay
    c <- c_start / (1:nIterations)^decay
    c[c<0.01] <- 0.01 #put a lower bound on c (dont make perturbations too small)
    #Generate initial parameters at random
    gammas <- runif(min=-0.05,max=0.05,n=p)
    betas  <- runif(min=-0.05,max=0.05,n=p)
    
    #Run iterations of SPSA
    for(i in 1:nIterations){
        #Generate perturbation vectors (bournoulli random variables of magnitude c[i])
        dGammas <- sample(x=c(-1,1),size=p,replace=TRUE)*c[i]
        dBetas  <- sample(x=c(-1,1),size=p,replace=TRUE)*c[i]
        
        #Compute the expectation values of the + and - configurations
        Fplus  <- ExpectationValue(n=n,edges=edges,p=p,gammas=gammas+dGammas,betas=betas+dBetas,nSamples=nSamples)
        Fminus <- ExpectationValue(n=n,edges=edges,p=p,gammas=gammas-dGammas,betas=betas-dBetas,nSamples=nSamples)
        
        #Compute the gradients of the parameters based on expectation values
        gGammas <- (Fplus - Fminus) / (2*dGammas)
        gBetas  <- (Fplus - Fminus) / (2*dBetas)
        
        #Update the parameters based on gradients
        gammas <- gammas + a[i] * gGammas
        betas  <- betas  + a[i] * gBetas
        
        #Report the progress
        print(paste('Iteration:',i,'Exp(+):',sprintf('%.4f',Fplus),'Exp(-):',sprintf('%.4f',Fminus)))
    }
}
\end{lstlisting}

The code can be run out of the box with the following example

\begin{lstlisting}
#The provided example problem
example <- function(){
    #4 qubits
    n <- 4
    #Edges of the max-cut problem in matrix from (rows are edges, column 1 is source vertex, column 2 is destination vertex)
    edges <- matrix(c( 0,1, 1,2 , 2,3 , 0,3),byrow=TRUE,ncol=2)
    #p=2 is sufficient for this problem
    p <- 2
    #A sufficient number of optimization iterations to solve problem
    nIterations <- 100
    #Typically need quite a few samples (measurements of quantum circuit) per iteration to 
    nSamples <- 10000
    #Heuristically chosen a and c
    a_start <- 0.25
    c_start <- 0.25
    decay <- 0.5
    SPSAforQAOA(n=n,edges=edges,p=p,nIteration=nIterations,nSamples=nSamples,a_start=a_start,c_start=c_start,decay=decay)
}
\end{lstlisting}

%% file: appendixqiskit.tex
\section{Complete code for python (Qiskit)}
\label{sec:qiskit}
Also available at \cite{code}
\begin{lstlisting}
from qiskit import *
from qiskit import QuantumCircuit, QuantumRegister, ClassicalRegister
from qiskit import execute
from qiskit.providers.aer import QasmSimulator

import random

#Create quantum circuit for QAOA from edges and parmeters
def QAOA(nQubits,edges,p,betas,gammas):  
    #Define quantum and classical registers
    qr = QuantumRegister(nQubits)
    cr = ClassicalRegister(nQubits)
    circuit = QuantumCircuit(qr,cr)
    
    #Initial Hadamards
    for q in range(nQubits):
        circuit.h(q)        
    #For the number of specified iterations
    for P in range(p):
        #Phase 1
        #For each edge
        for j in range(len(edges)):
            #First CNOT
            circuit.cx(edges[j][0],edges[j][1])           
            #Rz on target qubit
            circuit.rz(phi=gammas[P],qubit=edges[j][1])           
            #Second CNOT
            circuit.cx(edges[j][0],edges[j][1])
            
        for q in range(nQubits):
            circuit.rx(theta=2*betas[P],qubit=q)   
    circuit.measure(qr,cr)
    return circuit

#Return +1 or -1 with equal probability
def m1p1():
    return random.randrange(2)*2 - 1

#Compute all scores for a set of edges
def computeExpectationValue(counts,edges):
    totalScore = 0
    totalSamples = 0
    #For each bitstring measured (keys in counts dictionary)
    for bitstring in counts.keys():
        score = 0  #Score for this bitstring
        for j in range(len(edges)):
            if( bitstring[edges[j][0]] != bitstring[edges[j][1]] ):
                score = score + 1
        totalScore += score * counts[bitstring]  #Multiply score times the # of times it was observed
        totalSamples += counts[bitstring]        #Keep track of the number of measurements (samples)
    return(totalScore/totalSamples)

#Run the circuit and return counts
def runCKT(circuit,shots=10000,noise=True):
    simulator = Aer.get_backend('qasm_simulator')
    counts = execute(circuit,backend=simulator,shots=shots).result().get_counts(circuit)  
    return counts

#Run a circuit and get expectation value
def ExpectationValue(circuit,edges,nSamples=10000):
    #Run circuit and collect counts
    counts = runCKT(circuit=circuit,shots=nSamples)
    #Get the score of the counts
    score = computeExpectationValue(counts,edges)
    return(score)

def SPSAforQAOA(n,edges,p,nIterations,nSamples,a_start,c_start,decay):
    #Initiate
    a = []; c = [];
    for i in range(1,nIterations+1):
        a.append( a_start / (i ** decay) )
        c.append( c_start / (i ** decay) )
    for i in range(nIterations):
        if( c[i] < 0.01 ):
            c[i] = 0.01
            
    #Initiate gamma,beta
    gammas = []
    betas  = []
    for P in range(p):
        gammas.append( random.uniform(-.1,.1) )
        betas.append(  random.uniform(-.1,.1) )
    
    #Run iterations of SPSA
    for i in range(nIterations):
        #Randomly perturb gammas,betas by c[i]
        Delta_gammas = []; gammas_plus = []; gammas_minus = [];
        Delta_betas  = []; betas_plus  = []; betas_minus  = [];
        for P in range(p):
            #Generate perturbation vectors of bernoulli variables with magnitude c[i]
            Delta_gammas.append( m1p1() * c[i] )
            Delta_betas.append(  m1p1() * c[i] )
            #Create +/- versions of the parameters
            gammas_plus.append( gammas[P] + Delta_gammas[P])
            gammas_minus.append(gammas[P] - Delta_gammas[P])
            betas_plus.append( betas[P] + Delta_betas[P])
            betas_minus.append(betas[P] - Delta_betas[P])
         
        #Get the circuits for the +/- versions
        pCircuit = QAOA(nQubits=n,edges=edges,p=p,betas=betas_plus,gammas=gammas_plus)
        mCircuit = QAOA(nQubits=n,edges=edges,p=p,betas=betas_minus,gammas=gammas_minus)
        #Run the +/- versions
        Fplus =  ExpectationValue(circuit=pCircuit,edges=edges,nSamples=nSamples)  
        Fminus = ExpectationValue(circuit=mCircuit,edges=edges,nSamples=nSamples)  
        
        #Compute estimated gradients 
        g_gammas = []; g_betas = [];
        for P in range(p):
            g_gammas.append( (Fplus - Fminus) / (2*Delta_gammas[P]) )
            g_betas.append(  (Fplus - Fminus) / (2*Delta_betas[P])  )
            
        #Update the parameters
        for P in range(p):
            gammas[P] = gammas[P] + a[i]*g_gammas[P]
            betas[P]  = betas[P]  + a[i]*g_betas[P]
        
        #Report progress
        print('Iteration:',i,'Exp(+):',Fplus,'Exp(-):',Fminus)

def example():
    #4 qubits
    n = 4
    #Edges of the maxcut problem
    edges = [ [0,1] , [1,2] , [2,3] , [3,0] ]
    #p=2 is sufficient for this problem
    p = 2
    #A sufficient number of optimization iterations to solve problem
    nIterations = 100
    #Typically need quite a few samples (measurements of quantum circuit) per iteration to 
    nSamples = 10000
    #Heuristically chosen a and c
    a_start = 0.25
    c_start = 0.25
    decay = 0.5
    SPSAforQAOA(n=n,edges=edges,p=p,nIterations=nIterations,nSamples=nSamples,
    a_start=a_start,c_start=c_start,decay=decay)
\end{lstlisting}

%% file: appendixquantumops.tex
\section{Complete code for R (QuantumOps)}
\label{sec:quantumops}
Also available at \cite{code}
\begin{lstlisting}
library("QuantumOps")

#Create QAOA circuit (represented by a matrix) from edges and parameters
QAOA <- function(nQubits,edges,p,gammas,betas){
    #Initial Hadamard gates
    m <- many(gate=H(),n=nQubits)
    #For each stage
    for(P in 1:p){
        #Phase separation for each edge
        for(e in 1:dim(edges)[1]){
            m <- controlled(gate=X(),n=nQubits,cQubits=edges[e,1],tQubit=edges[e,2]) %*% m   #CNOT from control to target
            m <- single(gate=Rz(gammas[P]),n=nQubits,t=edges[e,2]) %*% m                     #Rz on target
            m <- controlled(gate=X(),n=nQubits,cQubits=edges[e,1],tQubit=edges[e,2]) %*% m   #CNOT from control to target
        }
        #Rx on all qubits
        m <- many(gate=Rx(2*betas[P]),n=nQubits) %*% m
    }
    m #Return the matrix
}

#classically check the score (cut) of a bitstring
computeScore <- function(bitstring,edges){
    score <- 0
    for(e in 1:dim(edges)[1]){  #For each edge  (R indices start at 1)
        if(bitstring[edges[e,1]+1] != bitstring[edges[e,2]+1]) #If bits on either end of the edge are different
            score <- score + 1                             #Score increases by 1
    }
    score #return score
}

#Find the expectation value of QAOA for a given set of parameters
ExpectationValue <- function(n,edges,p,gammas,betas,nSamples){
    #Generate the QAOA quantum circuit 
    circuit <- QAOA(nQubits=n,edges=edges,p=p,gammas=gammas,betas=betas)
    #Initialze a qubit register in the |00...0> state
    qr <- intket(x=0,n=n)
    #Apply the circuit (represented by a matrix) to the quantum register (represented by a vector)
    qr <- circuit %*% qr
    #Simulate the measurements of the quantum state (each measurement would correspond to individual run in experiment)
    measurements <- sample(x=0:(2^n-1),size=nSamples,prob=probs(qr),replace=TRUE) #Sample from 0 to n-1 with probabilities from the quantum state
    
    #Test the score of the measurements
    scoreSum <- 0     #Add up all the scores from all measurments
    for(i in 1:nSamples)  #convert integer result to binary bitstring before parsing
        scoreSum <- scoreSum + computeScore(bitstring=convert_dec2bin(measurements[i],len=n),edges=edges)
    expectationValue <- scoreSum/nSamples #Average them
    
    return(expectationValue)
}

#Run Experiment
SPSAforQAOA <- function(n,edges,p,nIterations,nSamples,a_start,c_start,decay=2){
    #Make learning rate and perturbation sequences
    a <- a_start / (1:nIterations)^decay
    c <- c_start / (1:nIterations)^decay
    c[c<0.01] <- 0.01 #put a lower bound on c (dont make perturbations too small)
    #Generate initial parameters at random
    gammas <- runif(min=-0.05,max=0.05,n=p)
    betas  <- runif(min=-0.05,max=0.05,n=p)
    
    #Run iterations of SPSA
    for(i in 1:nIterations){
        #Generate perturbation vectors (bournoulli random variables of magnitude c[i])
        dGammas <- sample(x=c(-1,1),size=p,replace=TRUE)*c[i]
        dBetas  <- sample(x=c(-1,1),size=p,replace=TRUE)*c[i]
        
        #Compute the expectation values of the + and - configurations
        Fplus  <- ExpectationValue(n=n,edges=edges,p=p,gammas=gammas+dGammas,betas=betas+dBetas,nSamples=nSamples)
        Fminus <- ExpectationValue(n=n,edges=edges,p=p,gammas=gammas-dGammas,betas=betas-dBetas,nSamples=nSamples)
        
        #Compute the gradients of the parameters based on expectation values
        gGammas <- (Fplus - Fminus) / (2*dGammas)
        gBetas  <- (Fplus - Fminus) / (2*dBetas)
        
        #Update the parameters based on gradients
        gammas <- gammas + a[i] * gGammas
        betas  <- betas  + a[i] * gBetas
        
        #Report the progress
        print(paste('Iteration:',i,'Exp(+):',sprintf('%.4f',Fplus),'Exp(-):',sprintf('%.4f',Fminus)))
    }
}

#The provided example problem
example <- function(){
    #4 qubits
    n <- 4
    #Edges of the max-cut problem in matrix from (rows are edges, column 1 is source vertex, column 2 is destination vertex)
    edges <- matrix(c( 0,1, 1,2 , 2,3 , 0,3),byrow=TRUE,ncol=2)
    #p=2 is sufficient for this problem
    p <- 2
    #A sufficient number of optimization iterations to solve problem
    nIterations <- 100
    #Typically need quite a few samples (measurements of quantum circuit) per iteration to 
    nSamples <- 10000
    #Heuristically chosen a and c
    a_start <- 0.25
    c_start <- 0.25
    decay <- 0.5
    SPSAforQAOA(n=n,edges=edges,p=p,nIteration=nIterations,nSamples=nSamples,a_start=a_start,c_start=c_start,decay=decay)
}
\end{lstlisting}